\documentclass[runningheads,a4paper]{llncs}

\usepackage{graphicx}
\usepackage{epstopdf}
\usepackage{amssymb}
\usepackage{listings}
\usepackage{multirow}
\usepackage{hhline}
\usepackage{rotating}
\usepackage{amsmath}
\usepackage[ruled,vlined]{algorithm2e}
\usepackage{subfig}

\usepackage{url}
\urldef{\mailsa}\path|{rlaishra, vvphoha}@syr.edu|    
\newcommand{\keywords}[1]{\par\addvspace\baselineskip
	\noindent\keywordname\enspace\ignorespaces#1}

\begin{document}
	
	\mainmatter  % start of an individual contribution
	
	% first the title is needed
	\title{Curie: A method for protecting SVM Classifier from Poisoning Attack}
	
	% a short form should be given in case it is too long for the running head
	\titlerunning{Curie: A method for protecting SVM Classifier from Poisoning Attack}
	
	% the name(s) of the author(s) follow(s) next
	%
	% NB: Chinese authors should write their first names(s) in front of
	% their surnames. This ensures that the names appear correctly in
	% the running heads and the author index.
	%
	\author{Ricky Laishram \and Vir Virander Phoha}
	\authorrunning{R. Laishram \and V. V. Phoha}
	% (feature abused for this document to repeat the title also on left hand pages)
	
	% the affiliations are given next; don't give your e-mail address
	% unless you accept that it will be published
	\institute{Department of Electrical Engineering and Computer Science\\
		Syracuse University, NY 13244-4100\\
		\mailsa\\}
	%\url{http://www.springer.com/lncs}}
	
	%
	% NB: a more complex sample for affiliations and the mapping to the
	% corresponding authors can be found in the file "llncs.dem"
	% (search for the string "\mainmatter" where a contribution starts).
	% "llncs.dem" accompanies the document class "llncs.cls".
	%
	
	%\toctitle{Lecture Notes in Computer Science}
	%\tocauthor{Authors' Instructions}
	\maketitle

	\begin{abstract}
		Machine learning is used in a number of security related applications such as biometric user authentication, speaker identification etc. A type of causative integrity attack against machine le arning called \emph{Poisoning attack} works by injecting specially crafted data points in the training data so as to increase the false positive rate of the classifier. In the context of the biometric authentication, this means that more intruders will be classified as valid user, and in case of speaker identification system, user A will be classified user B. In this paper, we examine poisoning attack against SVM and introduce - Curie - a method to protect the SVM classifier from the poisoning attack. The basic idea of our method is to identify the poisoned data points injected by the adversary and filter them out. Our method is light weight and can be easily integrated into existing systems. Experimental results show that it works very well in filtering out the poisoned data.
		
	\end{abstract}

	\section{Introduction}
	\label{sec:intro}
	Nowadays machine learning is used in a number of diverse applications. In the security domain, the use of machine learning techniques has become very important due to the emergence of big data. Machine learning techniques are widely used in areas such as biometric authentication \cite{kumar2016authenticating}, speaker identification \cite{greenberg2014nist}, malware detection in mobile platforms \cite{roy2015experimental} etc.
	
	In the security domain, the learning components are not static - they are continuously trained with new incoming data. This allows the classifier to adapt to changes in the system. For example, consider a biometric authentication system that uses machine learning to classify users as invalid or valid. The behavior of both the invalid and valid users will change over time. So the machine learning components need to be continuously updated to keep up with the new valid users. However, from a security perspective this opens up a new attack surface.
	
	In such applications, the environment that the system operates in is not completely under the control of the learner. There might be adversary\footnote{In this paper, the terms \emph{attacker} and \emph{adversary} are used interchangeably, and they refer to the same thing.} that actively tries to manipulate the classifier \cite{dalvi2004adversarial}. This creates a possible vulnerability in the security system because most of the machine learning algorithms were not originally developed to operate in such adversarial environments. 
	
	Most machine learning algorithms assume that the training and test data have the same distribution \cite{dalvi2004adversarial,bruckner2012static}. In the presence of an adversary, this assumption is no longer valid. The adversary can manipulate some of the training data to achieve some malicious goal - for example, to increase the false positive rate. In the case of the authentication system, this means that more invalid users will be classified as valid users.
	
	Barreno et al. \cite{barreno2006can} explored the poisoning attacks and broadly categorized them through two aspects. In terms of the influence that an attacker has over the training data, they classified the attack as causative and exploratory. In causative attacks, the attacker has knowledge of the classifier and some influence over the training data. In explorative attacks, the attacker can only probe the learner for information. In an online learning environment, the learner is susceptible to causative attacks. 
	
	In terms of the type of security violation, Barreno et al. classified the attacks as integrity attack and availability attack. In an integrity attack, the goal of the attacker is to increase the false positive rate. For example in an authentication system, an integrity attack would be an attempt to increase the rate of invalid users being classified as valid user.
	
	Over the years, various attack strategies have been developed against different learning algorithms. In this paper, we examine causative integrity attacks against a Support Vector Machine learner proposed by Biggio et al. \cite{biggio2012poisoning} and Xiao et al. \cite{xiao2012adversarial}. This is an important research topic because machine learning is used in a variety of environments where an adversary might be present, and SVM is one of the most widely used machine learning algorithms.
	
	The main contribution of this paper is a method, which we refer to as Curie\footnote{We called our method Curie from the word \emph{cure}, since the method is supposed to "cure" the classifier from the poisone data. It is also inspired by the character Curie in the game Fallout 4.}, to protect an SVM classifier against poisoned data injected by an adversary. Curie works by identifying the poisoned data points and filtering them out. To do this it relies on clustering the data in feature space, and then detecting the distances of the data points from other points in the same cluster in the (feature + label) space (Section \ref{sec:methodology-curie}). Experimental results (Section \ref{sec:results}) show that our proposed method can successfully identify most of the poisoned data. The method we propose is very light weight and can be easily integrated into any existing system.
	
	In the next section, we explore some background and related works. We provide a brief description of poisoning attacks against machine learning in Section \ref{sec:related-poison}. In Section \ref{sec:related-poison-svm}, we examine the poisoning attack against SVM proposed by Baggio et al. \cite{biggio2012poisoning} and Xiao et al. \cite{xiao2012adversarial}. In the next section (Section \ref{sec:related-solutions}), we examine some proposed solutions and their shortcomings. In Section \ref{sec:methodology}, we describe the threat model and our proposed solution. We describe the experimental setup and the dataset used in Section \ref{sec:experiment}. Section \ref{sec:results} presents the results of the experiments and some analysis. Finally, we present some concluding remarks and potential future works in Section \ref{sec:conclusion}. 
	
	\section{Background and Related Works}
	\label{sec:related}
	In this section, we will discuss some background related to poisoning attacks against machine learning algorithms. We will also discuss the poisoning attack againsts SVM proposed by Baggio et al. \cite{biggio2012poisoning} and Xiao \cite{xiao2012adversarial}, and some proposed solutions to protect against such attacks.
	
	\subsection{Poisoning Attack}
	\label{sec:related-poison}
	In many applications of machine learning in the security domain, the data is non-stationary, that is, the distribution of the data shifts over time. For example, in case, of a biometric authentication system, the features, such as walking pattern etc., for a valid user might change over time for a variety of reasons.
	
	\begin{figure}
		\centering
		\subfloat[Incremental update]{\includegraphics[width=0.45\textwidth]{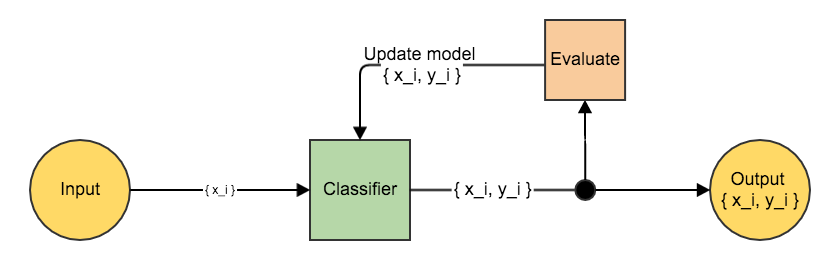}
			\label{fig:related-poison-1}}
		\hfill
		\subfloat[Periodic retraining]{\includegraphics[width=0.45\textwidth]{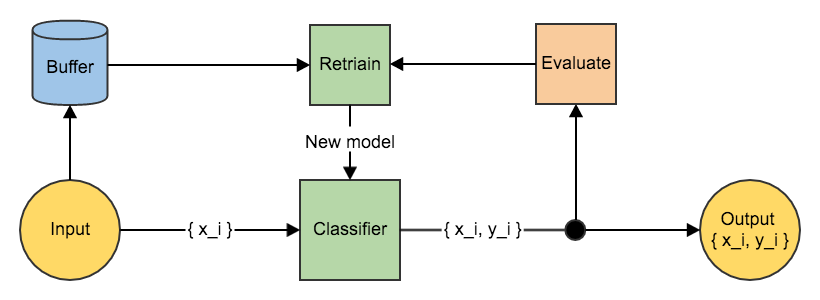}
			\label{fig:related-poison-2}}
		\caption{The models for (\ref{fig:related-poison-1}) incremental update and (\ref{fig:related-poison-2}) periodic retraining. In the incremental update model, whenever a new data is seen, the performance of the classifier on that data is evaluated and incremental update is made on the classifier. In the periodic retraining model, the data are stored in a buffer. When the performance of the classifier falls below a threshold or after a fixed time interval, the data stored in the buffer is used to retrain a new classifier.}
		\label{fig:related-poison-3}
	\end{figure}
	
	There are two ways to handle this non-stationary distribution - incremental algorithm or periodic retraining (Figure \ref{fig:related-poison-3}). In the first case, the model is updated incrementally as new data comes in, and in the second case, the data is buffered and the model is retrained periodically using the buffered data.
	
	\iffalse
	In the model with incremental update (Figure \ref{fig:related-poison-1}), an online classifier is used. There is an evaluator in the system that continuously checks the output from the classifier for classification errors. The classifier is then updated with the misclassified data if there is a miss-classification. In such system, there is no retraing phase.
	\fi
	
	In the model with periodic retraining (Figure \ref{fig:related-poison-2}), the inputs are stored in a buffer. The model is retrained after a fixed interval of time has passed or if the classifier performance falls below a pre-defined threshold. In our paper, this is the type of model that we are considering.
	
	Retraining the system opens up an attack surface \cite{barreno2006can,barreno2010security}. The data for retraining is collected from an adversarial environment. This opens up the possibility that an attacker could inject specially crafted data in the new training data with the goal of increasing the false positive rate. In the case of authentication system, this would mean that the retrained system will not be able to detect some invalid users. 
	
	In the next sub-section, we will discuss some proposed poison attack mechanisms against SVM classifier.
	
	\subsection{Poisoning Attack against SVM}
	\label{sec:related-poison-svm}
	In SVM, the fundamental idea is to find a decision surface   \iffalse \footnote{Although SVM classifiers can have non-linear decision surface, they are an extension of the linear classification thorough kernel trick. So, in this paper we are considering only linear classisfier.}\fi in the feature space that separates the training data based on the class labels \cite{cortes1995support}. To perform a poisoning attack against SVM, the attacker inserts some data in the training data with the labels flipped \cite{biggio2012poisoning,xiao2012adversarial}.
	
	In all the attack scenario described, the ability of the attacker is over-estimated. It is assumed that the attacker has complete knowledge of the classification system, and has the capability to insert some data into the retraining data.
	
	The naive way to perform the attack is to randomly flip the labels of the data to be used for retraining. This is equivalent to adding noise to the training data \cite{xiao2012adversarial}. Another way is to select the points that are near the support vectors and flip their labels. A third way is to select points that are farthest from the support vectors and flip the labels.
	
	Another method \cite{biggio2011support} is to select a combination of points that maximizes the classification error (or false positive rate). The problem with this approach, however, is that selection of the optimal set of such points is unfeasible for many real world data due to data size.
	
	More intelligent methods of selecting the data points to flip have been proposed. These methods \cite{xiao2012adversarial,biggio2012poisoning} works by selecting points that maximizes the loss function of the SVM classifier. In the work by Xiao et al. \cite{xiao2012adversarial}, a set of points is selected based on the loss maximization framework and their labels are flipped. In the work by Baggio et al.\cite{biggio2012poisoning}, some data points are selected and they are moved to other points in the feature space such that the loss is maximized. According to the results presented, both methods work very well in deteriorating the classification accuracy. In this paper, this is the type of attack we are addressing.
	
	In the next sub-section, we will describe some proposed solutions to protect a classsifier from poisoning attack.
	
	\subsection{Previous Methods to protect against Poisoning Attack}
	\label{sec:related-solutions}
	There are a few methods proposed for protection against poisoning attack. The first approach is to develop secure classifiers that are resistant to the attack. Most of the works \cite{bruckner2011stackelberg,bruckner2012static} on secure classifier algorithms uses a game theory model - the problem is modeled as a game between the adversary and the learner. The problem with this approach is that it is difficult to integrate with existing systems. Since our goal is to protect a system that uses SVM, this method is not of interest.
	
	Another method is to use multiple classifiers \cite{biggio2011bagging}. In the approach proposed, the poisoned data are treated as outliers and an ensemble of classifiers is used. However, a shortcoming of this method that they use from \(3\) to \(50\) base classifiers. The use of this many multiple classifiers is very resource intensive for learner. In addition, in the experiments performed in that work, they did not use the more intelligent methods \cite{biggio2012poisoning,xiao2012adversarial} of generating the poison data.
	
	A third approach is to hide information about the classifier from the adversary. However, as mentioned in Section \ref{sec:related-poison-svm}, we are considering the worst case scenario where he attacker has complete knowledge of the classifier. So, we will not consider this approach.
	
	In this paper, we consider the poisoning attack with gradient ascent proposed by Biggio and Xiao \cite{biggio2012poisoning,xiao2012adversarial}. As far as we know, there has been no work done the SVM classifier against such attacks. In the next section, we will describe the threat model and our methodology to protect SVM against such attack.
	
	\section{Methodology}
	\label{sec:methodology}
	In Section \ref{sec:related-poison-svm}, we described various poisoning attack strategies. \iffalse The first method of random label flipping is equivalent to adding noise to the data and can be handled as such. The second method of selecting points close to the decision boundary will not have a significant impact as demonstrated by Xiao et al \cite{xiao2012adversarial}. The third method of selecting points farthest form the decision boundary can be solved by considering those points as outliers. \fi In this paper, we are addressing the type of attack \cite{xiao2012adversarial,biggio2012poisoning} in which the attack points are selected to maximize the loss function. We propose a method to protect an SVM classifier from this type of poison attack. Our goal is to develop a method that can identify the attack points added by the adversary, and filter them out before the data is used for retraining.
	
	In this section we will discuss the threat model first. Then, we will describe our proposed method in detail.
	
	\subsection{Threat Model}
	\label{sec:methodology-threat}
	In this paper, the type of attack we are taking into consideration is a causative integrity attack - the adversary has complete knowledge of the system and is trying to increase the false positive rate of the system. In the real world, the adversary is not likely to possess complete knowledge about the classifier. So, the adversary would perform exploratory attacks first. In our case, we make the assumption that the adversary has complete knowledge of the classifier for simplicity \footnote{By complete knowledge we mean that the adversary knows everything about the classifier - from the training data to the hyper-parameters used.}.
	
	Assume that there is a classifier that uses SVM. For simplicity, assume that we are dealing with a two class classification - positive and negative classes. In the attack scenario considered in this paper, the attacker has the capability to inject some data into the training data. In the experiments presented is Section \ref{sec:experiment}, we consider different cases according to the amount of data that the attacker has added.
	
	The attacker injects as few specially crafted data points as possible so that the false positive rate is increased. So, the objective of the attack is to increase the number of negative class misclassified as positive, without significantly changing the number of positive class misclassified as negative. The attack points are generated using the method proposed by Biggio et al. \cite{biggio2012poisoning}. The attack method proposed by Xioa et al. \cite{xiao2012adversarial} is very similar. So, we consider only one attack method in this paper. 
	
	\subsection{Curie - Poison Points Filter}
	\label{sec:methodology-curie}
\begin{figure}
		\centering
		\includegraphics[width=0.75\textwidth]{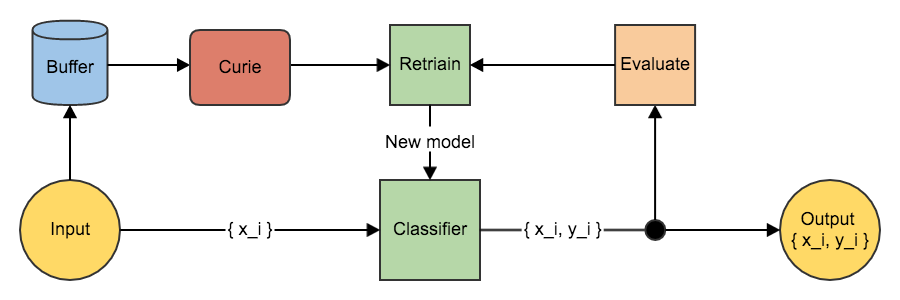}
		\caption{The figure shows the integration of Curie with the periodic retraining model. The only difference between this and Figure \ref{fig:related-poison-3} is the Curie module between Buffer and Retrain modules in \ref{fig:methodology-1}. Curie filters out the attack points before retraining the model.}
		\label{fig:methodology-1}
	\end{figure}

	The core idea of our proposed method, Curie, lies in the identification of the data points that the attacker has added. We will refer to this data as poison points. Once these poison points are identified, they can be filtered out before the data is used for retraining the classifier. Figure \ref{fig:methodology-1} shows how our method fits in with the existing periodic retraining model.
	
	In the model with periodic retraining (Figure \ref{fig:related-poison-2}) described in Section \ref{sec:related-poison}, the attacker injects the poison points in the buffer. So, in our method, we add a filter between the buffer and the retrain module as shown in Figure \ref{fig:methodology-1}.
	
	As mentioned in Section \ref{sec:related-poison-svm}, the attack works by selecting a set of points in the feature space that maximizes the loss function, and then flipping their labels. An additional constrain for the attacker is that these points should be well hidden within the rest of the valid points.
	
	\begin{figure}
		\centering
		\subfloat[Clusters in feature space]{\includegraphics[width=0.45\textwidth]{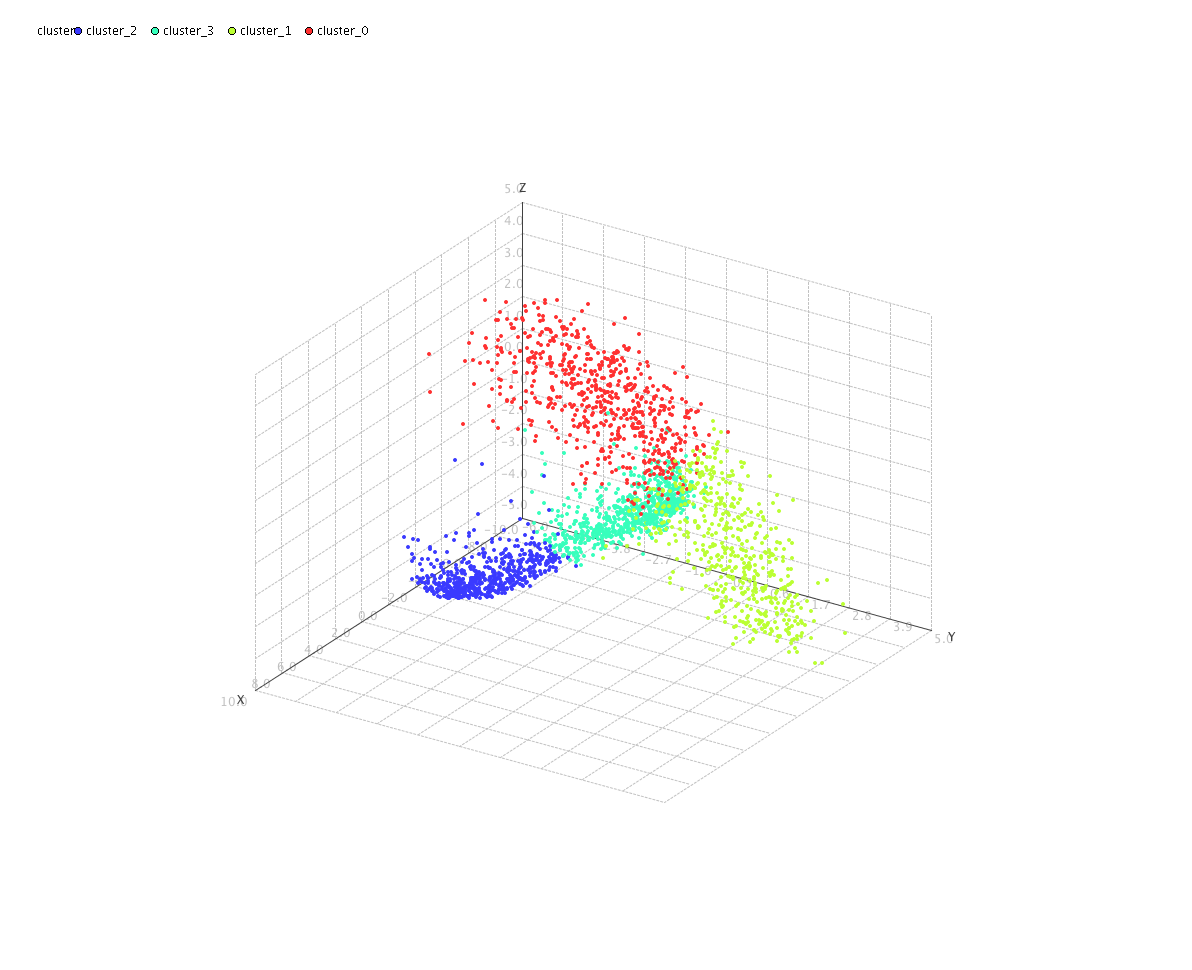}
			\label{fig:methodology-2}}
		\hfill
		\subfloat[Clusters in (feature + label) space]{\includegraphics[width=0.45\textwidth]{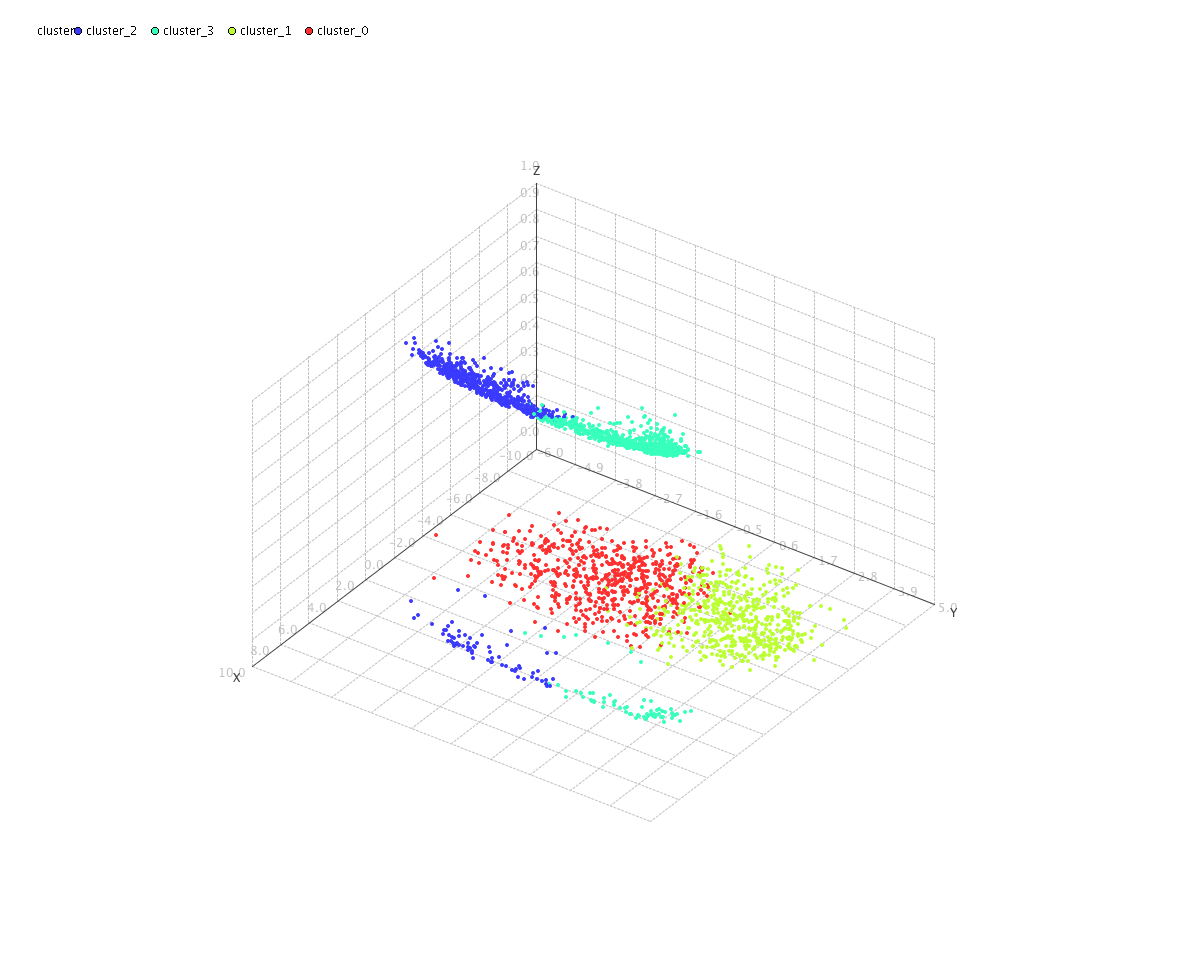}
			\label{fig:methodology-3}}
		
		\caption{Clusters in feature space (Figure \ref{fig:methodology-2}) and (feature + label) space. The different clusters are represented by different color and the clustering algorithm used is DBSCAN. In Figure \ref{fig:methodology-2}, the axes are the first three principal components. In Figure \ref{fig:methodology-3}, the x and y axes are the first two principal components and the z-axis is the class label.}
		\label{fig:methodology-4}
	\end{figure}
	
	Figure \ref{fig:methodology-2} shows the clusters in the feature space after the poison points have been injected. It can be observed that there are some outliers but experiments (Section \ref{sec:experiment}) shows that these are not the poison points.
	
	However, we can make use of the label flipping to distinguish the poison points from the normal points. Because of the fact that labels of the poison points have been flipped, the poison points will be very similar to the normal points in the feature space. 
	
	In our approach, we extend the feature space to an additional dimension which maps the class labebls to some values. We will refer to this additional dimension as the \emph{label dimnension} and the new feature space will be refered to as \emph{(feature + label) space}.
	
	In the (feature + label) space, the poison points will be well seperated from the rest of the cluster in feature space if the mapping of the label dimension is done properly as shown in \ref{fig:methodology-3}. (We will talk about how to do the mapping in the later part of this section.)
	
	From these observations, we develop the attack points identification method as described in Algorithm \ref{alg:1}.
	
	\iffalse
	The first step in our method is to reduce the number of feature dimensions though Principal Component Analysis. In our algorithm, we select the first \(n\) principal components that accounts for \(0.95\) of the variance in the data. If the number of dimensions is not very high, this step can be skipped.
	\fi
	
	The first step in our method is to cluster the data in the feature space. In our algorithm, the clustering algorithm used is DBSCAN \cite{ester1996density}. In our algorithm, DBSCAN is used because it can find clusters of arbitrary shape and we do not need to specify the number of clusters.
	
	As mentioned earlier, we are dealing with a two class classification. Assume that the two classes are \(\{0, 1\}\). It does not matter which class is \(0\) or \(1\). Let \(class(x)\) be the class label of \(x\). The mapping in the label dimension is done such that of \(class(x)\) is mapped to \(\omega \cdot class(x)\).
	
	Then a point \(x\) in the (feature + label) space is given by,
	
	\[F(x) = \{x_1, x_2, \ldots , \omega \cdot class(x) \}\]
	
	Let us assume that the cluster to which \(x\) belongs to is given by \(cluster(x)\). Let \(S\) represent the entire set of data points. Let \(S^\prime(x)\) represent the set of points that belongs to the same cluster as \(x\).
	
	\[S^\prime(x) = \{y \vert y \in S \wedge Cluster(x) = Cluster(y) \}\]
	
	Compute the average squared Euclidean distance for each point in \(S^\prime(x)\) from \(x\). Let us call this the distance score for \(x\) and denote it by \(d(x)\).
	\[d(x) = \frac{1}{\vert S^\prime(x) \vert} \cdot \sum_{y \in S^\prime(x)}{dist(x,y)}\]
	
	where, \(dist(x,y)\) is the squared Euclidean distance between \(x\) and \(y\).
	
	Let \(D\) be the set of distance score for all points in the dataset.
	\[D = \{d(y) \vert y \in S\}\]
	
	Assume \(\mu\) and \(\sigma\) are the mean and standard deviation of \(D\). Let \(Z(x)\) represent the z-score of a data point \(x\).
	
	\[Z(x) = \frac{x - \mu}{\sigma}\]
	
	Transform every point in \(D\) to its z-score. Let \(\overline{D}\) denote this new set.
	\[\overline{D} = \{Z(x) \vert \forall  x \in D\} \]
	
	Filter out the points with confidence less than \(conf(\theta)\),
	\[S^* = \{y \vert y \in S \wedge \overline{D}(y) \le \theta  \} \]
	
	The set \(S^*\) is the training data from which the poisoned data has been removed, and it is the output of Curie which will be used for future retaining of the SVM classifier. 
	
	In the algorithm described, there are two hyperparameters - \(\omega\) and \(\theta\). In Appendix \ref{ap:curie-hyperparameters} we show that,
	
	\[\omega ^2 \ge \frac{\theta^2 -1}{(1-\rho)} \cdot icd_{max}\]
	
	where \(icd_{max}\) is the maximum intra-cluster distance and \(\rho\) is the expected probability that a data point will be a poison point.
	
	In the next subsection, we will describe an extension of Curie to take in account multi-class classification.
	
	\begin{algorithm}[!t]
		\SetAlgoLined
		\KwData{$Data = (F, C)$ such that $F$ is a set of feature vectors and $C$ is a set of class labels}
		\KwResult{$M$ vectors after removal of attack points}
		\tcc{Reduce the number of dimensions}
		$PcaData \gets$ PCA($Data.F$) \;
		\tcc{Cluster the data thorough DBSCAN}
		$Clusters \gets$ DBSCAN($PcaData$)\;
		\ForEach{$point \in Data$}{
			\tcc{Append the weighted class as feature}
			$point.F \gets$ Append($point.F, point.C \times \omega $)\;
			\tcc{Calculate the average distance to $10$ randomly selected points in the same cluster}
			$cls \gets$ GetCluster($point, Clusters$)\;
			$sample \gets$ Sample($cls,count$)\;
			\ForEach{$s \in sample$}{
				$s.F \gets$ Append($s.F, s.C \times weight $)\;
				$d \gets$ EucledianDistance($point.F, s.F$) \;
				$Dist.point \gets Dist.point + d$ \;
			}
			$Dist.point \gets Dist.point/$Size($cls$)\; 
		}
		\tcc{Perform Z-score standarization and select only points with more than $conf(\theta)$ confidence}
		$Dist \gets$ ZScore($Dist$)\;
		\ForEach{$point \in Data$}{
			\If{$Dist.point \le \theta$}{
				$Result \gets $ Append($Result, point$)\;
			}
		}
		\caption{Curie: Algorithm to filter the poison points}
		\label{alg:1}
	\end{algorithm}
	
	\subsection{Curie in a Multi-Class Classification System}
	\label{sec:methodology-curie-multi}
	In the previous sections, we considered only two-class classification. This is valid for authentication systems as they only need to classify users as valid or invalid. However, there are security systems in which the classifier has to handle multiple classes. In this section, we present an extension to Curie for multi-class classification.
	
	Consider a multi-class classifier with \(N_C\) number of classes. From the perspective of the attacker, this is not different from the two-class classification. If the class of interest to the attacker is \(C_A\), the attack can be performed by considering the classifier as binary classifier with classes - \(C_A\) and the rest. 
	
	Since Curie is an unsupervised algorithm, it can be easily extended to apply to such cases of multi-class classification. The defender does not know the class that the attacker is targeting. So, we consider every possible one-vs-rest pairs. This means that instead of adding one label dimension in the (feature + label) space like in Section \ref{sec:methodology-curie}, \((N_C - 1)\) additional dimensions are added in this case. The rest of the algorithm remains the same. To differentiate this from the two-class version, we will refer to this as MultiClass-Curie. As shown in Appendix \ref{app:m-curie-hyper}, the equation for \(\omega\) given by \ref{eq:curie-hyper-10} will apply for MultiClass-Curie as well.
	
	In the next section, we will describe the dataset used for experiments, and experimental setups.
	
	\section{Datasets and Experiments}
	\label{sec:experiment}
	In this section, we will first describe the datasets that we are using for our experiments. Then we will describe the experimental setups for the three different experiments.
	
	\subsection{Datasets}
	\label{sec:experiment-data}
	To verify our method, we perform experiments using the MNIST dataset \cite{lecun1998mnist}. The MNIST dataset consist of \(10\) classes and aproximately \(60000\) images of dimensions \(28\times28\). We convert each of the images into a vector of length \(784\) where each value represents a pixel in the image. Each item of the vector has a value in the range \([0,255]\).
	
	As described in Section \ref{sec:experimental-setup}, we need two datasets - one with \(2\) classes and one with \(3\) classes.
	
	To create the \(2\) class dataset, we randomly sample \(1250\) points without replacement from classes \(0\) and \(1\) in the original dataset. For the \(3\) class dataset, we randomly sample \(1500\) data points without replacement from classes \(0\),  \(1\) and \(2\). So the first dataset is of size \(2500\) and the second is \(4500\).
	
	Then we generate poison points for each of these two datasets. We use the method proposed by Biggio et al. \cite{biggio2012poisoning} using the AdversariaLib library. For each of the datasets, we generate \(25\), \(50\), \(75\), \(100\) and \(125\) poison points to create \(5\) additional instances of each dataset.
	
	\subsection{Experimental Setup}
	\label{sec:experimental-setup}
	For our experiments, we use a linear SVM classifier, with penalty (\(C\)) of value \(1\). We do not implement the entire system shown in Figure \ref{fig:methodology-1} since the purpose of Curie is only to find the poison points.
	
	We have three separate experimental setups. The first experiment is to determine the effectiveness of Curie in removing the poison points compared to an outlier detection algorithm. The second experiment is to explore the effect of changes in the performance of Curie due to change in the hyperparameters. The third experiment is to determine the effectiveness of MultiClass-Curie.
	
	As mentioned in Section \ref{sec:related-solutions}, we are not aware of any method to protect SVM classifiers against the attacks proposed in \cite{biggio2012poisoning,xiao2012adversarial}. So, we evaluate the performance of Curie by comparing it to the case without Curie.
	
	\subsubsection{Experiment 1}
	As mentioned in the Section \ref{sec:experiment-data}, we have five instances of the \(2\) class MNIST dataset based on the amount of poison data injected, and one instance without any poison points. 
	
	In this experiment, the filtered data from Curie is used to an train an SVM classifier and we compare the changes compared to the case when the training data is used for training directly. We perform 10-fold cross validation and user the accuracy and false positive rate for comparison.
	
	\subsubsection{Experiment 2}
	In the second experiment, we fixed the threshold \((\theta)\) at \(1.645\) and change the values of the weight \((\omega)\) to observe the effect on the performance of Curie.
	
	We use the two class dataset described in Section \ref{sec:experiment-data} and inject \(2\%\) poison points. For this part of this experiment, we vary \(\omega\) from \(1\) to \(10^4\). We perform 10-fold cross validation and calculate the classifier accuracy and false positive rate for each value of \(\omega\).
	
	\subsubsection{Experiment 3}
	In this experiment, we use the three class data described in Section \ref{sec:experiment-data}. Since SVM does not natively support multi-class classification, we use the one-vs-rest approach. We train the classifier on the different instance of this dataset, and calculate the accuracy 10-fold cross validation. 
	
	The training data is passed through MultiClass-Curie and the classification is performed again. We then compare the change in the classifier performance. We are not aware of any other works on preventing poisoning attacks in multi-class SVM classifier. So, we evaluate the performance of our algorithm by the amount of improvement over the case with no protection.
	
	In the next section, we will present the results of the experiments described here and some analysis.
	
	\section{Results and Analysis}
	\label{sec:results}
	
	In the previous section we described the MNIST dataset and the experimental setup. In this section, we present the results of the experiments described is Section \ref{sec:experimental-setup}.
	
	\subsection{Results of Experiment 1}
\begin{figure}
		\centering
		\subfloat[Accuracy comparison]{\includegraphics[width=0.45\textwidth]{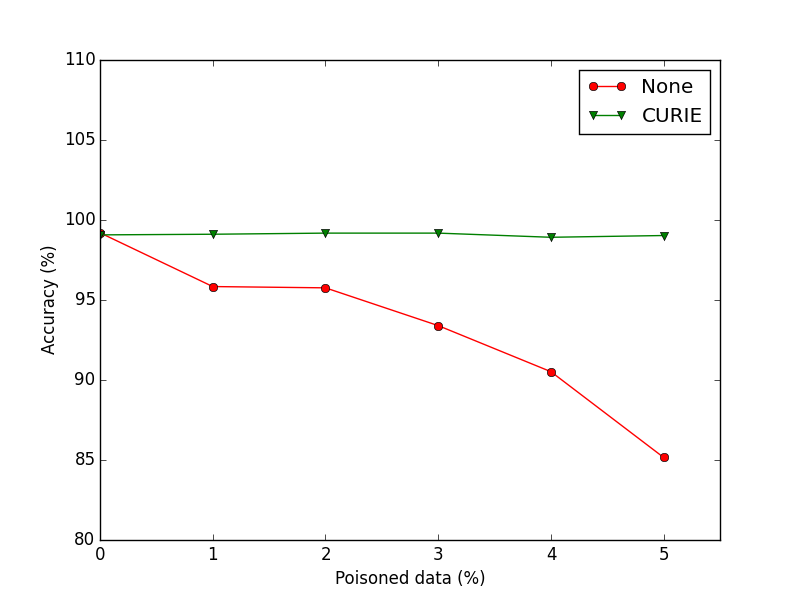}
			\label{fig:result-1}}
		\hfill
		\subfloat[False Positive Rate comparison]{\includegraphics[width=0.45\textwidth]{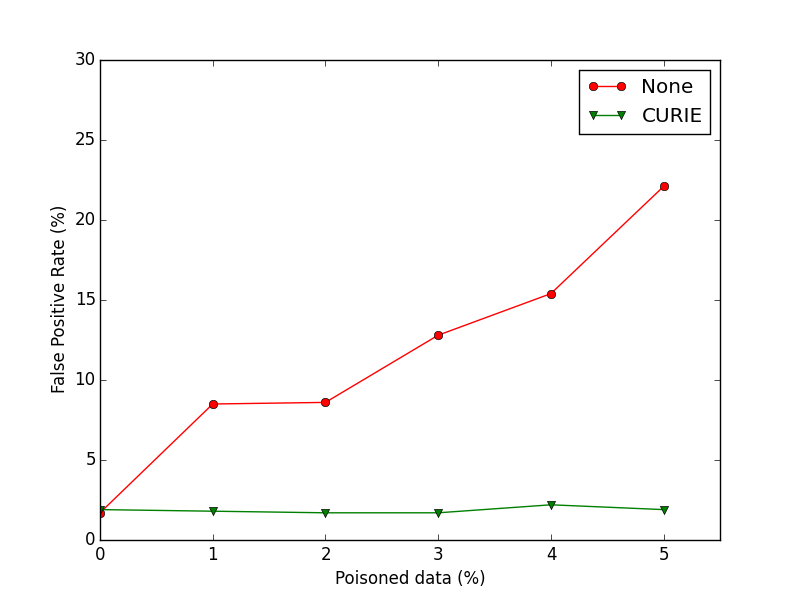}
			\label{fig:result-2}}
		\caption{Comparison between the performance of the SVM classifier with and without Curie. Figure \ref{fig:result-1} is the plot of the accuracy vs percentage of poison points and Figure \ref{fig:result-2} is the plot of the False Positive Rate vs percentage of poison points. In both plots, the green line represents the performance of the SVM classifier with Curie and the red line is the baseline without Curie.}
		\label{fig:result-3}
	\end{figure}

	The performance comparison of the SVM classifier with and without Curie is given in Table \ref{tab:result-1}. Figure \ref{fig:result-1} and \ref{fig:result-2} presents a plot of the accuracy and false positive rate for various amount of poison points injected.
	
	\begin{table}
		\centering
		\caption{Accuracy and False Positive Rate comparison with and without Curie for different amount of poison points injected.}
		\begin{tabular}{| c c | c c | c c|}
			\hline
			\multicolumn{2}{|c}{Poison Points} & \multicolumn{2}{c}{None} & \multicolumn{2}{c|}{Curie} \\ \hhline{------}
			Number & \% & Accuracy & FPR & Accuracy & FPR \\ \hline
			0 & 0 & 0.992 & 0.017  & 0.990 & 0.019 \\
			25 & 1 & 0.958 & 0.085  & 0.991 & 0.018 \\
			50 & 2 & 0.957 & 0.086  & 0.991 & 0.017 \\
			75 & 3 & 0.934 & 0.128  & 0.991 & 0.017 \\
			100 & 4 & 0.905 & 0.154  & 0.989 & 0.022 \\
			125 & 5 & 0.851 & 0.221  & 0.990 & 0.019 \\ \hline
		\end{tabular}
		\label{tab:result-1}
	\end{table}
	
	In the plots in Figure \ref{fig:result-3}, the green line represents our method - Curie. In both Figure \ref{fig:result-1} and \ref{fig:result-2}, it can be observed that with Curie the accuracy and false positive rate of the classifier remains almost constant even when the amount of poison points are increased. This indicates that it is filtering out the poison points before they are used to train the SVM classifier. These results shows that Curie is very effective in identifying and removing poison points regardless of the amount of poison points injected.
	
	\subsection{Results of Experiment 2}
	In this section, we present the results of Experiment 2. For this experiment we used only the instance of the \(2\) class dataset with \(50\) poison points.
	
	\begin{table}
		\centering
		\caption{False Positive Rate and Accuracy comparison for various values of \(\omega\).}
		\begin{tabular}{| c | c | c |}
			\hline
			\(\omega\) & False Positive Rate & Accuracy \\
			\hline
			1 & 0.108 & 0.948 \\
			10 & 0.082 & 0.961 \\
			100 & 0.025 & 0.990 \\
			1000 & 0.015 & 0.992 \\
			3000 & 0.015 & 0.992 \\
			5000 & 0.015 & 0.992 \\
			10000 & 0.150 & 0.917 \\ \hline
		\end{tabular}
		\label{tab:result-2}
	\end{table}
	
	\begin{figure}
		\centering
		\subfloat[Accuracy comparison]{\includegraphics[width=0.45\textwidth]{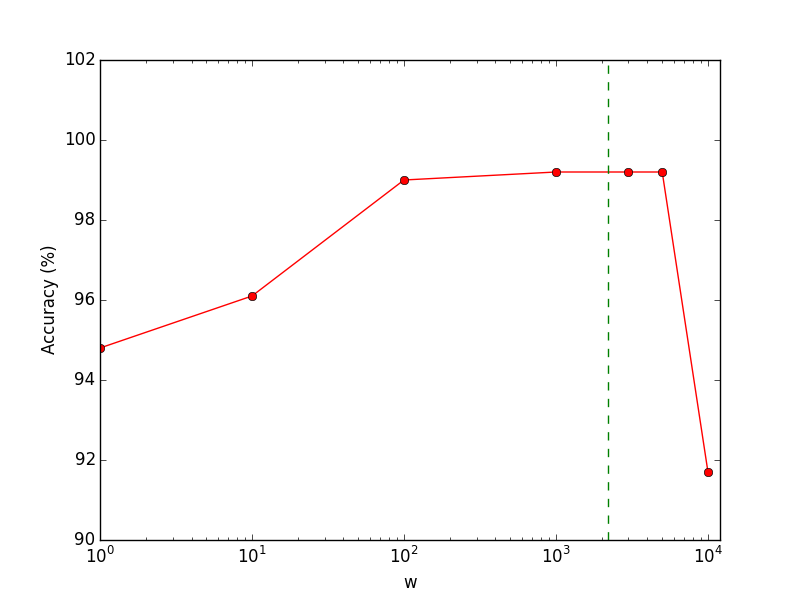}
			\label{fig:result-4}}
		\hfill
		\subfloat[False Positive Rate comparison]{\includegraphics[width=0.45\textwidth]{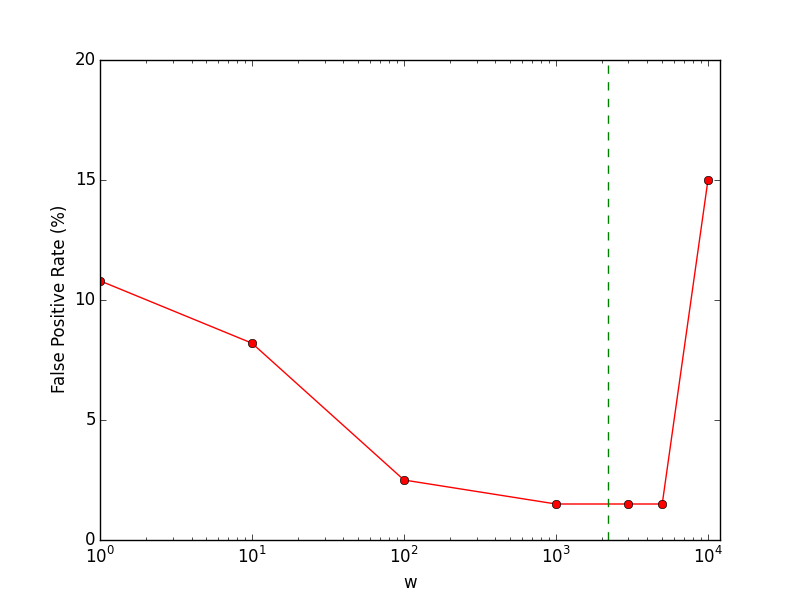}
			\label{fig:result-5}}
		\caption{Comparison between the performance of Curie for various values of \(\omega\). Figure \ref{fig:result-4} shows the accuracy and Figure \ref{fig:result-5} shows the false positive rate for various values of \(\omega\). (Note that the scale of the x-axis is logarithmic.)}
		\label{fig:result-6}
	\end{figure}
	
	In Figure \ref{fig:result-4}, the accuracy of the classifier for various values of \(\omega\) is shown and Figure \ref{fig:result-5} shows the false positive rate of the classifier. The vertical green dashed line represents the theoretical value of \(\omega\) given by \ref{eq:curie-hyper-10} with \(\rho = 0.02\). 
	
	These results shows that Curie is not very sensitive to \(\omega\), and in turn \(\rho\) (Equation \ref{eq:curie-hyper-10}). In a real world dataset, the exact value of \(\rho\) in unknown. So, the results shows that Curie can work in such cases. 
	
	\subsection{Results of Experiment 3}
	In this subsection, we present the results of the third experiment. The accuracy comparison of the classifier with and without MultiClass-Curie for different amount of poison points is given in Table \ref{tab:result-3}.
	
	\begin{table}
		\centering
		\caption{Accuracy comparison of the classifier with and without MultiClass-Curie for different amount of poison points.}
		\begin{tabular}{| c c | c c |}
			\hline
			\multicolumn{2}{|c}{Poison Points} & \multicolumn{2}{c|}{Accuracy} \\ \hhline{----}
			Number & \%  & None & Curie \\ \hline
			0 & 0 & 0.978 & 0.976 \\
			25 & 0.55 & 0.975 & 0.979 \\
			50 & 1.11 & 0.968 & 0.975 \\
			75 & 1.67 & 0.957 & 0.970 \\
			100 & 2.22 & 0.933 & 0.965 \\
			125 & 2.78 & 0.918 & 0.961 \\ \hline
		\end{tabular}
		\label{tab:result-3}
	\end{table}
	
	\begin{figure}
		\centering
		\subfloat[Accuracy comparison]{\includegraphics[width=0.45\textwidth]{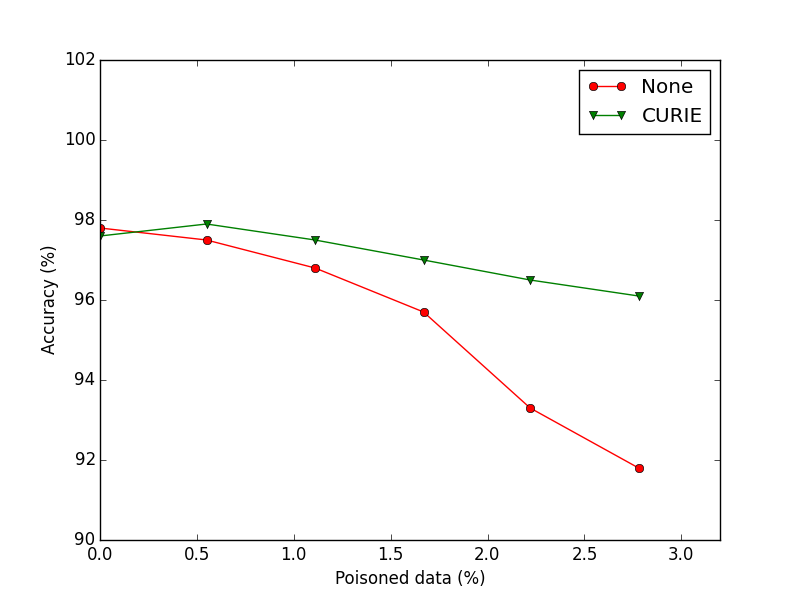}
			\label{fig:result-7}}
		\caption{Accuracy comparison between the multi-class classifier performance with and without MultiClass-Curie. The red line represents the accuracy of the classifier without MultiClass-Curie, and the green line represents the accuracy with MultiClass-Curie.}
		\label{fig:result-8}
	\end{figure}
	
	Figure \ref{fig:result-8} shows the comparison of the classifier accuracy for different amount of poison points with and without MultiClass-Curie. It can be observed that on its own the accuracy of the classifier decreases as the number of the poison points increases. However, with MultiClass-Curie, the accuracy decreases very slowly with the number of poison points. These results shows that in the case of multi-class classifier, MultiClass-Curie can still be used to protect the SVM classifier from poisoning attacks.
	
	\section{Conclusion}
	\label{sec:conclusion}
	The use of machine learning for security applications is becoming very widespread. So, an investigation of the behavior of machine learning algorithms in an adversarial environment is important. Poisoning attacks present a very real threat to security systems (Section \ref{sec:related}) that employ machine learning techniques. Researches \cite{biggio2012poisoning,xiao2012adversarial} have shown that SVM, a very widely used machine learning algorithm, is very susceptible to poisoning attacks.
	
	In this paper, we examined the poisoning attack against SVM, and propose a method, which we refer to as Curie, to protect a system that uses SVM classifier from such attacks. Curie works as a filter before the buffered data are used to retrain the SVM classifier. 
	
	Curie works by exploiting the fact that the poison data looks like a normal point in the feature space, but has their labels flipped. The data are clustered in the feature space, and the average distance of each point from the other points in the same cluster is calculated with the class label considered as a feature with proper weight. The data points with confidence less than \(95\%\) confidence are removed from the training data.
	
	We tested our algorithm on the MNSIT dataset (Section \ref{sec:experiment-data}), and evaluated with the case in which the training data is used directly for retraining. The results obtained (Section \ref{sec:results}) show that when Curie is used, the accuracy and false positive rate of the classifier does not change significantly when the amount of poison points in the training data is changed. This indicates that Curie successfully filtered out most of the poison data. We also showed that Curie is not very sensitive to \(\omega\). So the estimation of \(\rho\) for an unknown dataset need not be very accurate. We also demonstrated experimentally that we can extend Curie to make it work in a multi-class classification system.
	
	In addition to the performance benifits, Curie can be easily integrated into any system that uses periodic retraining (Figure \ref{fig:methodology-1}).
	
	Since Curie is an unsupervised method, an attacker would have to use evasion attacks to pass through Curie. This means that the attacker would have to inject data that works both for evasion attack and poison attack. (The injected data point has to pass thorough Curie undetected, and then poison the classifier.) We are not aware of any works on attacks that does both both evasion and poisoning simultaneously. An interesting research area would be to explore if this is possible.
	
	Although we tested Curie with only SVM classifier, it should be possible to extend the work to protect other classifiers, such as perceptron, against poisoning attack. Like in the case of SVM, perceptron also works by creating a decision boundary. So the poisoning attack against perceptron would also work through some form of label flipping. As we demonstrated in Section \ref{sec:methodology-curie}, label flipping creates poison points that are similar to normal points in feature space but different in the (feature + label) space. This is what Curie exploits to filter out the poison points.
	
	\bibliography{references.bib}
	\bibliographystyle{splncs03}
	
	\appendix
	\section{Curie Hyperparameters}
	\label{ap:curie-hyperparameters}
	In Section \ref{sec:methodology-curie}, we described the algorithm for Curie. In the algorithm described, we use two hyperpareters - \(\theta\) and \(\omega\). In this section, we will describe these hyperparemets and the relation between them.
	
	In Curie, the hyperparameter \(\omega\) is the weight used for mapping the the class label to the additional dimension in the (feature + label) space. The parameter \(\theta\) defines the threshold above which the a data point in the training data is considered to be a poison point.

	Assume that the positive class is mapped to \(x_+\) and the negative class to \(x_-\) in the additional dimension introduced in the (feature + label) space. We define \(\omega\) in our algorithm (Section \ref{sec:methodology-curie}) such that,
	
	\begin{equation}
	\label{eqn:curie-hyper-1}
	\omega \ge	\vert x_+ - x_- \vert
	\end{equation}
	
	Consider a cluster \(C\) of size \(n\) in the feature space. Let \(S_c\) be the set of data points in cluster \(C\). Suppose that there are \(l\) dimensions in the data originally. 
	
	As mentioned in Section \ref{sec:methodology-curie}, the distance scores are computed only for pairs of points that belong to the same cluster. Then, in if we consider the cluster \(C\) in the (feature + label) space, for \(x_+^c, x_-^c \in S_c\), let \(\omega_c\) be,
	
		\begin{equation}
		\label{eqn:curie-hyper-1.b}
		\omega_c \ge \vert x_+^c - x_-^c\vert
		\end{equation}
	
	Since Equation \ref{eqn:curie-hyper-1} is describes \(\omega\) over the entire dataset,
	
		\begin{equation}
		\label{eqn:curie-hyper-1.c}
		\omega \ge	\omega_c
		\end{equation}
	
	Consider a poison point \(p\) in cluster \(C\). The average distance\footnote{The distance metric used is the squared euclidean distance.} of \(p\) from the rest of the points in cluster \(C\) in the feature space is given by,
	
	\begin{equation}
		\label{eqn:curie-hyper-2}
		\overline{dist}_f(p) = \frac{1}{n - 1} \sum_{x \in S_c - \{p\}}\left(\sum_{i = 1}^{l}{(x_i - p_i)^2}\right)
	\end{equation}
	
	Consider a data point \(d\) in the cluster \(C\). Assume \(d_m\) is the value of the additional dimention of \(d\) in (feature + label) space. If \(p\) is a poison point in cluster \(C\), the label of \(p\) was flipped. So, 
	
	\[d_m = d_+ \implies p_m = p_-\]
	\[d_m = d_- \implies p_m = p_+\]
	
	In the (feature + label) space, the average distance of \(p\) from the rest of the points in \(C\) is given by,
	
	\begin{equation}
		\label{eq:curie-hyper-3}
		\begin{aligned}
			\overline{dist}_{f+l}(p) &= \frac{1}{n - 1} \sum_{x \in S_c - \{p\}}\left(\sum_{i = 1}^{l + 1}{(x_i - p_i)^2}\right) \\
			&= \frac{1}{n - 1} \sum_{x \in S_c - \{p\}}\left( (x_m - p_m)^2 + \sum_{i = 1}^{l}{(x_i - p_i)^2}\right)
		\end{aligned}
	\end{equation}
	
	For the algorithm to be able to distinguish \(p\) in the (feature + label) space from non-poison data, the average distance of the poison points in the (feature + label) space should be significantly greater than that in the feature space. That is,
	
	\begin{equation}
		\label{eq:curie-hyper-4}
		\theta \cdot \overline{dist}_f(p) \le \overline{dist}_{f+l}(p)
	\end{equation}
	
	Substituting Equations \ref{eqn:curie-hyper-2} and \ref{eq:curie-hyper-3} in Equation \ref{eq:curie-hyper-4}, we get,
	
	\begin{equation}
		\label{eq:curie-hyper-5}
		\begin{aligned}
			\frac{\theta}{n - 1} \sum_{x \in S_c - \{p\}}\left(\sum_{i = 1}^{l}{(x_i - p_i)^2}\right) &\le \frac{1}{n - 1} \sum_{x \in S_c - \{p\}}\left( (x_m - p_m)^2 + \sum_{i = 1}^{l}{(x_i - p_i)^2}\right) \\
			\theta \sum_{x \in S_c - \{p\}}\sum_{i = 1}^{l}{(x_i - p_i)^2} &\le \sum_{x \in S_c - \{p\}} \left( (x_m - p_m)^2 + \sum_{i = 1}^{l}{(x_i - p_i)^2} \right)\\
			(\theta - 1) \sum_{x \in S_c - \{p\}}\sum_{i = 1}^{l}{(x_i - p_i)^2} &\le \sum_{x \in S_c - \{p\}} (x_m - p_m)^2
		\end{aligned}
	\end{equation}
	
	Consider the sum on the right hand side of Equation \ref{eq:curie-hyper-5}. For an \(x \in S - \{p\} \), if \(x\) were also a poison point, \(x_m = p_m\). Then, in this case, \((x_m-p_m)^2 = 0\). In case \(x\) is not a poison point,
	
	\[(x_m - p_m)^2 \le \omega_c^2\]
	
	Assume that there is a probability \(\rho\) of a point being a poison point. The number of normal points in \(S_c - \{p\}\) is \((1 - \rho)(n - 1)\). Then,
	
	\begin{equation}
		\label{eq:curie-hyper-6}
		(\theta -1 ) \sum_{x \in S_c - {p}}\sum_{i = 1}^{l}{(x_i - p_i)^2} \le (1-\rho) (n-1) \omega_c^2
	\end{equation}
	
	\begin{equation}
		\label{eq:curie-hyper-7}
		\omega_c ^2 \ge \frac{\theta^2 -1}{(1-\rho)(n-1)} \sum_{x \in S_c - {p}}\sum_{i = 1}^{l}{(x_i - p_i)^2}
	\end{equation}
	
	In the feature space, the poison point indistinguishable a normal point. So, the average intra-cluster distance of cluster \(C\) can be approximated by,
	
	\begin{equation}
		\label{eq:curie-hyper-8}
		icd(C) \approx \frac{1}{(n-1)} \sum_{x \in S_c - {p}}\sum_{i = 1}^{l}{(x_i - p_i)^2}
	\end{equation}
	
	Then Equation \ref{eq:curie-hyper-7} can be written as,
	
	\begin{equation}
		\label{eq:curie-hyper-9}
		\omega_c ^2 \ge \frac{\theta -1}{(1-\rho)} \cdot icd(C)
	\end{equation}
	
	If there are \(m\) clusters, let us use \(\mathbb{C}\) to represent the set of all clusters. Then from Equation \ref{eqn:curie-hyper-1.c},
	
	\begin{equation}
	\label{eq:curie-hyper-11}
	\omega \ge max(\{\omega_i \vert i \in \mathbb{C} \})
	\end{equation}
	
	The value \(\theta\) and \(\rho\) is defined over the entire dataset. Let us use \(icd_{max}\) to denote the maximum value of intra cluster distance over all the clusters in \(\mathbb{C}\).
	
	\[ icd_{max} = max(\{icd(i) \vert i \in \mathbb{C}\})\]
	
	Then the overall value of \(\omega\) for the entire dataset is,
	\begin{equation}
	\label{eq:curie-hyper-10}
	\omega^2 \ge \frac{\theta -1}{(1-\rho)} \cdot icd_{max}
	\end{equation}
	
	\section{MultiClass-Curie Hyperparameters}
	\label{app:m-curie-hyper}
	In this Section, we will show that the equation for \(\omega\) given by Equation \ref{eq:curie-hyper-10} is valid for MultiClass-Curie as well.
	
	As mentioned in Section \ref{sec:methodology-curie-multi}, in MultiClass-Curie there are \((N_C-1)\) additional dimensions in the (feature + label) space. So, Equation \ref{eq:curie-hyper-3} will be rewritten as,
	
	\begin{equation}
		\label{eq:m-curie-hyper-1}
		\begin{aligned}
			\overline{dist}_{f+l}(p) &= \frac{1}{n - 1} \sum_{x \in S_c - {p}}\left(\sum_{i = 1}^{l + N_C - 1}{(x_i - p_i)^2}\right) \\
			&= \frac{1}{n - 1} \sum_{x \in S_c - {p}}\left(\sum_{i = 1}^{l}{(x_i - p_i)^2} + \sum_{j=0}^{N_C-1}(x_j - p_j)^2\right)
		\end{aligned}
	\end{equation}
	
	As mentioned in Section \ref{sec:methodology-curie-multi}, when the attacker construct the attack, the classifier is considered as a two classifier - one class representing the class of interest \(C_A\) and the other class representing the rest of the classes. So, the difference \(x_j - p_j\) will be \(0\) in all cases except the one that corresponds to \(C_A\) vs the rest. That is,
	
	\[\sum_{j=0}^{N_C-1}(x_j - p_j)^2 = (x_m - p_m)^2\]
	
	So, Equation \ref{eq:m-curie-hyper-1} can be written as,
	
	\begin{equation}
		\label{eq:m-curie-hyper-2}
		\overline{dist}_{f+l}(p) = \frac{1}{n - 1} \sum_{x \in S_c - {p}}\left((x_m - p_m)^2 + \sum_{i = 1}^{l}{(x_i - p_i)^2}\right)
	\end{equation}
	
	Equation \ref{eq:m-curie-hyper-2} is identical to \ref{eq:curie-hyper-3} and the number of classes does not come up anywhere else in the proof in Appendix \ref{ap:curie-hyperparameters}. So, the rest of the proof is identical and Equation \ref{eq:curie-hyper-10} will give the value of \(\omega\) for MultiClass-Curie as well.
	
	\iffalse
	\section{Results for MNIST-0-1 dataset}
	\label{ap:mnist-table}
	\begin{sidewaystable}[!t]
		\centering
		\caption{Complete Results for MNIST-0-1 dataset}
		\begin{tabular}{| c | c | c | c | c | c | c | c | c | c |}
			\hline
			Algorithm& Poison Points Count & Poison Points (\%) & TP& FP & TN & FN & FPR & FNR & Accuracy(\%) \\ \hline	
			None&0&	0&		1326&	20&	1154&	0&		0.017&	0&		99.20 \\
			&25&	2&		1300&	103&	1096&	1&		0.085&	0&		95.84\\
			&50&	4&		1276&	106&	1118&	0&		0.086&	0&		95.76\\
			&75&	6&		1247&	161&	1088&	4&		0.128&	0.003&		93.40\\
			&100&	8&		1186&	197&	1077&	40&		0.154&	0.032&		90.52\\
			&125&	10&		1118&	288&	1011&	83&		0.221&	0.069&		85.16\\ \hline	
			LDCOF &0&	0&		1236&	20&	1154&	0&		0.017&	0&		99.17\\
			&25&	2&		1213&	88&	1111&	0&		0.073&	0&		96.35\\
			&50&	4&		1188&	104&	1120&	0&		0.084&	0&		95.69\\
			&75&	6&		1164&	144&	1105&	0&		0.115&	0&		94.03\\
			&100&	8&		1040&	179&	1095&	1&		0.14&	0&		92.55\\
			&125&	10&		1116&	215&	1084&	0&		0.165&	0&		91.10\\ \hline
			Curie & 0 &	0 &		1299 &	23 &	1151 &	0	&	0.019 &	0	&	99.07\\
			&25	&2	&	1299&	21&	1153&	1&		0.018&	0&		99.11\\
			&50	&4	&	1276&	20&	1154&	0&		0.017&	0&		99.18\\
			&75	&6	&	1251&	20&	1154&	0&		0.017&	0&		99.18\\
			&100&	8&		1226&	26&	1148&	0&		0.022&	0&		98.92\\
			&125&	10&		1201&	23&	1151&	0&		0.019&	0&		99.03\\ \hline
		\end{tabular}
	\end{sidewaystable}
	\fi
	
\end{document}